\documentclass[conference]{IEEEtran}
\IEEEoverridecommandlockouts
\usepackage{cite}
\usepackage{amsmath,amssymb,amsfonts}
\usepackage{algorithmic}
\usepackage{graphicx}
\usepackage{textcomp}
\usepackage{xcolor}
\usepackage{hyperref}
\usepackage{physics}
\usepackage{subcaption}
\usepackage{svg}

\usepackage{fancyhdr}

\def\BibTeX{{\rm B\kern-.05em{\sc i\kern-.025em b}\kern-.08em
    T\kern-.1667em\lower.7ex\hbox{E}\kern-.125emX}}

\makeatletter
\renewcommand{\IEEEauthorrefmark}[1]{\textsuperscript{#1}}
\makeatother

\begin{document}

\title{Platform-Agnostic Modular Architecture for Quantum Benchmarking

\thanks{This work was done in collaboration with the Quantum Economic Development Consortium (QED-C) and was performed under the auspices of the QED-C Technical Advisory Committee on Standards and Performance Metrics. The authors acknowledge many committee members for their input and feedback on the project and this manuscript.}
}

\author{
  \IEEEauthorblockN{
    Neer Patel,\IEEEauthorrefmark{1} 
    Anish Giri,\IEEEauthorrefmark{2, 3}
    Hrushikesh Pramod Patil,\IEEEauthorrefmark{2, 4}
    Noah Siekierski,\IEEEauthorrefmark{9} 
    Avimita Chatterjee,\IEEEauthorrefmark{5}
    Sonika Johri,\IEEEauthorrefmark{6}\\
    Timothy Proctor,\IEEEauthorrefmark{9}
    Thomas Lubinski,\IEEEauthorrefmark{7, 8}
    Siyuan Niu\IEEEauthorrefmark{1}
  }
\\
\IEEEauthorblockA{\IEEEauthorrefmark{1}\textit{University of Central Florida, Orlando, FL 32816, USA}}
\IEEEauthorblockA{\IEEEauthorrefmark{2}\textit{QED-C -- SRI International, Arlington, VA 22209, USA}}
\IEEEauthorblockA{\IEEEauthorrefmark{3}\textit{University of Illinois Urbana-Champaign, Champaign, IL 61820, USA}}
\IEEEauthorblockA{\IEEEauthorrefmark{4}\textit{North Carolina State University, Raleigh, NC 27606, USA}}
\IEEEauthorblockA{\IEEEauthorrefmark{5}\textit{Computer Science \& Engineering, Pennsylvania State University, State College, PA 16801, USA}}
\IEEEauthorblockA{\IEEEauthorrefmark{6}\textit{Coherent Computing Inc, Cupertino, CA, USA}}
\IEEEauthorblockA{\IEEEauthorrefmark{7}\textit{QED-C Technical Advisory Committee -- Standards, Arlington, VA 22209, USA}}
\IEEEauthorblockA{\IEEEauthorrefmark{8}\textit{Quantum Circuits Inc, New Haven, CT 06511, USA}}
\IEEEauthorblockA{\IEEEauthorrefmark{9}\textit{Sandia National Laboratories, Albuquerque, NM 87123, USA}}
}

\maketitle

\begin{abstract}

We present a platform-agnostic modular architecture that addresses the increasingly fragmented landscape of quantum computing benchmarking by decoupling problem generation, circuit execution, and results analysis into independent, interoperable components.
Supporting over 20 benchmark variants ranging from simple algorithmic tests like Bernstein-Vazirani to complex Hamiltonian simulation with observable calculations,
the system integrates with multiple circuit generation APIs (Qiskit, CUDA-Q, Cirq) and enables diverse workflows. We validate the architecture through successful integration with Sandia's \textit{pyGSTi} for advanced circuit analysis and CUDA-Q for multi-GPU HPC simulations.
Extensibility of the system is demonstrated by implementing dynamic circuit variants of existing benchmarks and a new quantum reinforcement learning benchmark, which become readily available across multiple execution and analysis modes.
Our primary contribution is identifying and formalizing modular interfaces that enable interoperability between incompatible benchmarking frameworks, demonstrating that standardized interfaces reduce ecosystem fragmentation while preserving optimization flexibility.
This architecture has been developed as a key enhancement to the continually evolving QED-C Application-Oriented Performance Benchmarks for Quantum Computing suite.
\end{abstract}



\pagestyle{fancy}

\renewcommand{\headrulewidth}{0.0pt}
\lhead{}
\rhead{\thepage}

\renewcommand{\footrulewidth}{0.4pt}
\cfoot{}

\lfoot{Platform-Agnostic Modular Architecture for Quantum Benchmarking }

\rfoot{\today}

\section{Introduction}
\label{sec:introduction}

Quantum computing benchmarking \cite{Proctor2025-cd} has evolved rapidly in recent years, in parallel to significant advances in quantum hardware capabilities, algorithmic development, and software toolchains. We have seen the emergence of increasingly sophisticated methodologies that assess performance of quantum systems from low-level gate fidelities to high-level application performance~\cite{cross2019validating, qiskit_measuring_quantum_volume, baldwin2022re, pelofske2022quantum,proctor2022measuring, blume2020volumetric, proctor2022establishing, kharkov2022arline, Supermarq, QASMbench, Quark, nation2025benchmarking, PracticalCharacter}. The breadth and depth of these benchmarking efforts are representative of the overall maturing of the quantum computing field. Rigorous performance evaluation is highly important as quantum systems make a transition from research prototypes to practical computational tools.

However, the proliferation of different benchmarking approaches creates challenges for the quantum computing community. Benchmarking efforts span multiple levels of the quantum computing stack, including component-level hardware characterization~\cite{Proctor2025-cd, PracticalCharacter}, system-level performance evaluation\cite{QASMbench}, measurement of compiler and toolchain  \cite{kharkov2022arline}, and algorithmic/application-oriented performance analysis ~\cite{Supermarq, lubinski2023_10061574}.  Each of these comes with its own unique methodology and metrics and provides valuable insights within its focus area. The resulting landscape presents users with dozens of independent benchmarking projects, each claiming distinct advantages and targeting different aspects of quantum system evaluation. This fragmentation creates substantial barriers for researchers and practitioners who lack the time and resources to evaluate multiple benchmarking frameworks, leading to confusion about which approaches are most suitable for specific evaluation needs.

Our work addresses some of these concerns with a modular architecture that unifies benchmarking approaches across algorithm and application levels while providing integration pathways for system-level and tooling-level evaluation frameworks. We do not claim this to be a complete solution to the fragmentation problem. Yet, our architecture represents a step toward reducing barriers between different benchmarking methodologies and enabling more comprehensive quantum system evaluation. This enhanced architecture incorporates integration mechanisms that enable integration with third-party tooling that offers complementary execution pathways and analytics. 

\begin{figure}[t!]
\centering
\includegraphics[width=\linewidth]{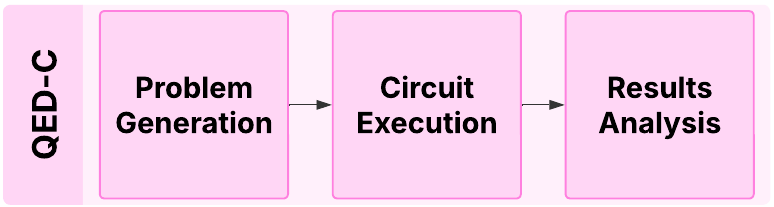}
\caption{\textbf{QED-C Framework Modularization.} The new QED-C architecture modularizes benchmarks into three components: problem generation, circuit execution, and results analysis. Each module consists of easy-to-access methods or variables for utilization.} 
\label{fig: QEDC_Architecture}
\end{figure}

The key architectural contribution is the decomposition of the benchmarking process into three independent and interoperable components: problem generation, circuit execution, and results analysis, as shown in Figure~\ref{fig: QEDC_Architecture}. We focus on gate-model quantum computing because it is capable of universal quantum computation and is the most widely adopted paradigm in current quantum hardware. This modular design enables users to leverage any combination of these components according to their specific needs, whether they utilize the complete integrated suite or integrate individual components with external tools. The architecture supports multiple circuit generation APIs and provides standardized interfaces for custom execution backends, enabling integration with specialized hardware systems, compiler optimizations, and third-party analysis frameworks. This architecture has been developed as a key enhancement to the continually evolving suite of QED-C Application-Oriented Performance
Benchmarks for Quantum Computing~\cite{lubinski2023_10061574
,lubinski2023optimization,lubinski2024quantumalgorithmexplorationusing,chatterjee2024comprehensivecrossmodelframeworkbenchmarking,niu2025practicalframeworkassessingperformance}.

We demonstrate this flexibility through implementations that integrate with established quantum computing tools, including \textit{pyGSTi}~\cite{noauthor_undated-bl, Nielsen2020-rd} for advanced circuit analysis and CUDA-Q~\cite{CUDA-Q} for multi-GPU HPC simulations~\cite{MPI,slysz2025hybridclassicalquantumsupercomputingdemonstration,guo2025qgearimprovingquantumsimulation,xu2024atlashierarchicalpartitioningquantum,zhong2025scalableparallelsimulationquantum}. These integrations are validated through the execution of several QED-C benchmarks on backend quantum computing systems, such as IBM Quantum \cite{ibmq2025} and NERSC Perlmutter NVIDIA GPU simulators~\cite{Perlmutter}. Additionally, we demonstrate the architecture's extensibility by implementing new benchmark variants, including dynamic circuit instances and a Quantum Reinforcement Learning (QRL)~\cite{qaio25, kruse2025cleanqrl} benchmark, which become readily available across these execution and analysis modes, as part of the benchmark problem collection.

\section{Background}
\label{sec:background}

\subsection{QED-C Application-Oriented Benchmarks}\label{subsec:QC-App-Oriented}
\label{sec:application_oriented_benchmarks}

The Quantum Economic Development Consortium (QED-C) has developed an open-source benchmark suite that builds upon prior years of quantum benchmarking methodologies~\cite{PhysRevA.77.012307, PhysRevLett.106.180504, Blume-Kohout2017-no, Cross_2019, Boixo_2018, wack_clops_2021}.
Unlike system-level benchmarks such as Quantum Volume (QV)~\cite{cross2019validating, qiskit_measuring_quantum_volume, baldwin2022re, pelofske2022quantum} and Volumetric Benchmarking (VB)~\cite{proctor2022measuring, blume2020volumetric, proctor2022establishing}, the QED-C suite provides application-specific programs that enable comprehensive performance assessment across various quantum hardware and simulators, including high-performance NVIDIA GPUs~\cite{lubinski2023_10061574, lubinski2023optimization}.
QED-C suite measures execution quality, runtime costs, and resource requirements for both single-circuit executions and iterative algorithms like QAOA~\cite{farhi2014quantum} and VQE~\cite{peruzzo2014variational}, using normalized Hellinger fidelity to assess circuit quality under noise~\cite{lubinski2023_10061574}. 

The QED-C benchmark framework has supported multiple quantum programming APIs (Qiskit, Cirq, AWS Braket) since inception, providing access to diverse simulation and hardware backends. The recent integration with CUDA-Q introduces ``quantum kernels" that embed classical computation within quantum circuits, a structure that has influenced the framework enhancements described in this manuscript.


\subsection{Complementary Benchmarking Frameworks}

Numerous quantum benchmarking libraries and frameworks have been developed, including Supermarq \cite{Supermarq}, Benchpress \cite{nation2025benchmarking}, QASMBench \cite{QASMbench}, Quark\cite{Quark}, Arline benchmark \cite{kharkov2022arline}, etc. In this work, we choose \emph{pyGSTi} for integration with QED‑C due to its complementary functionality to the QED‑C benchmarking suite, which is detailed below.

\textbf{\emph{pyGSTi}} is an open-source Python package developed by Sandia's Quantum Performance Laboratory (QPL) that implements a broad suite of methods for evaluating noisy quantum computing devices \cite{noauthor_undated-bl, Nielsen2020-rd}. It contains implementations of a wide variety of characterization, benchmarking, and noise simulation methods. Two capabilities of \emph{pyGSTi} are particularly relevant to the QED-C benchmarking suite (and other benchmarking libraries), as described below.

First, \emph{pyGSTi} contains tools for creating scalable and efficient benchmarks from any quantum circuit or algorithm \cite{proctor2022measuring, proctor2022establishing}. Integrating the QED-C benchmarking suite with \emph{pyGSTi} would enable any QED-C benchmark to be transformed into a scalable and robust test for quantifying a system’s performance on that circuit. Second, \emph{pyGSTi} can generate complex and realistic error models \cite{Blume-Kohout2022-ln, Sarovar2020-pz, Proctor2020-iz} that capture, for example, coherent errors \cite{Blume-Kohout2022-ln}, crosstalk errors \cite{Sarovar2020-pz}, and time-varying errors \cite{Proctor2020-iz}, and can simulate circuits under these conditions, including efficient simulation for certain noise models and circuit types \cite{Miller2025-xp}. This enables testing and analysis of the QED-C benchmarking suite under realistic noise scenarios, as demonstrated in this work.


\subsection{High-Performance Simulation with CUDA-Q}

As quantum computing systems grow in scale, speed, and execution fidelity, they bring the promise of large-scale quantum applications that offer practical quantum advantage~\cite{huang2025vastworldquantumadvantage,
aaronson2025futurequantumcomputing,
Herrmann_2023}. In the near-term, classical high-performance GPU-accelerated (HPC) quantum simulation provides both speed and computational power in addition to high-fidelity execution~\cite{ramesh2025benchmarking,faj2023quantumcomputersimulationswarp,chandani2024efficientchargepreservingexcitedstate,ma2024understandingestimatingexecutiontime,Gangapuram_2024,montanezbarrera2025evaluatingperformancequantumprocessing}. As such, HPC quantum simulation can provide a platform for the exploration and validation of quantum applications that would execute at larger scales on next-generation hardware systems~\cite{Miessen_2024,dente2013gpuacceleratedtrottersuzukisolver,WITTEK2015339,KAWASE2023108720}.

In HPC applications that make use of NVIDIA CUDA or CUDA-Q, it is common to use MPI~\cite{MPI} to execute portions of the application across multiple GPUs to reduce the total execution time and to expand the overall memory footprint of the application~\cite{ramesh2025benchmarking,faj2023quantumcomputersimulationswarp,chandani2024efficientchargepreservingexcitedstate,ma2024understandingestimatingexecutiontime}. In the implementation of quantum applications requiring similar techniques, MPI can increase the state space of the simulation by stitching together the states across multiple processes, thereby increasing the effective number of qubits in the simulation.

In prior work, we demonstrated a prototype integration of the CUDA-Q programming and simulation API with the QED-C application-oriented benchmarking suite~\cite{Chatterjee_2025,niu2025practicalframeworkassessingperformance,ramesh2025benchmarking}.
This initial work involved cloning and modifying code for several benchmarks to use CUDA-Q and to execute on NVIDIA GPU quantum simulators, including Quantum Fourier Transform, Phase Estimation, and two additionial benchmarks. Recent efforts expanded this approach to full-scale quantum Hamiltonian simulation using Hamiltonians extracted from HamLib~\cite{hamlib2023}. These benchmarks evaluated both the speed and quality of Trotterized Hamiltonian circuits as well as the performance of observable computation. Through this prototype implementation, we identified several opportunities for architectural improvements to enhance code efficiency, scalability, and API flexibility, which we address in detail later in this manuscript.


\subsection{Dynamic Circuits}

\begin{figure}[t!]
  \centering
  \begin{subfigure}[ht]{0.50\columnwidth}
    \centering
    \includegraphics[width=\linewidth]{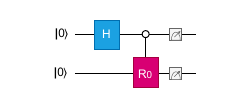}
    \caption{Static circuit}
    \label{fig:sub:a}
  \end{subfigure}%
  \hfill%
  \begin{subfigure}[ht]{0.50\columnwidth}
    \centering
    \includegraphics[width=\linewidth]{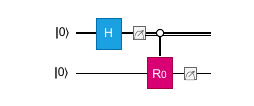}
    \caption{Dynamic circuit}
    \label{fig:sub:b}
  \end{subfigure}

  \caption{%
    \textbf{Static vs. Dynamic Circuit.}
    (a) In the static circuit, the Hadamard and controlled-\(R_{0}\) gates are applied before the final measurement.
    (b) In the dynamic variant, the first qubit is measured immediately after the Hadamard, and the classical output is used to conditionally trigger the single qubit \(R_{0}\) rotation. Finally, the second qubit is measured. 
  }
  \label{fig: static Vs dynamic}
\end{figure}

A dynamic quantum circuit allows measurements to be performed in the middle of the circuit and uses the outcomes to control subsequent quantum operations in real time. Unlike static circuits, where all gates are fixed in advance and measurements are performed only at the end, dynamic circuits enable conditional logic and adaptive control based on mid-circuit results.
Figure~\ref{fig: static Vs dynamic} illustrates a simple example that shows the difference between a static and a dynamic circuit. 
In the dynamic version, the first qubit is measured immediately after the Hadamard gate, and the result determines in real time whether to apply the single-qubit rotation $R_{0}$ to the second qubit before its measurement. In the static version, the Hadamard and two-qubit rotations are applied in a fixed sequence, with all measurements performed at the end.

Mid-circuit measurement (MCM) with Reset can enable qubit reuse, which reduces the circuit width and number of gates, thereby lowering the overall resource overhead~\cite{Fang2023DynamicCompile, decross2023qubitreuse, hua2023caqr, niu2024effective}. When combined with real-time feed-forward, MCM truly defines a dynamic circuit. It is not only a key component for achieving quantum error correction in fault-tolerant quantum computing, but also creates new opportunities for resource optimization.

For example, implementing an $n$-qubit Quantum Fourier Transform (QFT) followed by measurement can be reduced from $O(n^2)$ two-qubit gates in a static circuit to $O(n)$ mid-circuit measurements in a dynamic version, without connectivity constraints~\cite{Baumer2024DynamicQFT}. Similarly, Córcoles \emph{et al.} demonstrated that adaptive iterative phase estimation with MCM achieves lower error rates and requires fewer measurements than Kitaev’s static approach~\cite{Corcoles2021DynamicQPE}. Dynamic circuits have also enabled more efficient state and unitary preparation, achieving constant or logarithmic depth with fewer multi-qubit gates compared to static methods~\cite{Niu2024ACDC, buhrman2023state, hashim2024efficient, baumer2025measurement, smith2023deterministic, liao2025achieving, alam2024learning}. These algorithmic advances are supported by recent progress in quantum hardware that enables MCM and real-time feed-forward control, with platforms such as IBM, Quantinuum, IonQ, and QuEra, paving the way for practical applications of dynamic circuits.

\subsection{Quantum Reinforcement Learning}\label{subsec:Arch}
 
Machine learning has rapidly become essential across numerous industries. Advances in quantum computing have further enhanced machine learning, giving rise to the emerging field of Quantum Machine Learning (QML) \cite{biamonte2017quantum}. Carefully designed QML methods demonstrate potential advantages over classical methods; Abbas et al. experimentally illustrated superior expressivity and trainability on IBM's 27-qubit device \cite{abbas2021power}, while Saggio et al. highlighted benefits specifically within quantum reinforcement learning settings \cite{saggio2021experimental}.

Reinforcement Learning (RL) is a foundational paradigm in machine learning that unifies optimal control theory, Markov decision processes (MDPs), and adaptive decision-making \cite{kaelbling1996reinforcement}. In RL, an agent interacts with an environment $\mathcal{E}$, and given a state $s_t$ selects actions $a_t$ according to a policy $\pi(a_t|s_t)$, and receives scalar rewards $R_t$ as feedback. The central objective is to identify an optimal policy $\pi^*$ that maximizes the expected cumulative reward. This objective is formalized through the value function $V^\pi(s)$, defined by the Bellman equation \cite{Krylov1980}:

\begin{equation}
V^\pi(s) = \mathbb{E}_\pi \left[ R_{t+1} + \gamma V^\pi(S_{t+1}) \mid S_t = s \right].
\end{equation}

\begin{figure}[t!]
\centering
\includegraphics[width=\linewidth]{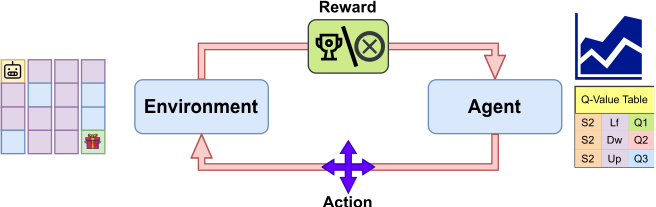}
\caption{\textbf{Illustration of Reinforcement Learning.} Figure showing the interaction between agent and environment, along with policy and value functions (e.g., Q-value table).}
\label{fig:QRL_basic}
\end{figure}

As depicted in Figure~\ref{fig:QRL_basic}, RL comprises two core entities: the agent and the environment. The environment may be instantiated as a software program or as a physical system, while the agent may correspond to either a software implementation or a real-world robotic system. The agent observes the environment, selects actions according to a policy or value function, and aims to maximize the rewards provided by the environment. Many paradigms exist within reinforcement learning, including actor–critic models \cite{konda_actor_critic}, Q-learning \cite{watkins1992q}, and proximal policy optimization \cite{schulman2017proximal}. With the advent of deep learning, traditional and often costly RL approaches have been significantly augmented through the integration of deep neural architectures \cite{sutton2018reinforcement}, enabling scalability to high-dimensional state and action spaces\cite{DeepQNet}.

Deep Reinforcement Learning (Deep RL) addresses the scalability challenges of classical RL. For instance, Q-learning \cite{watkins1992q} maintains explicit Q-value tables, which become intractable in high-dimensional or continuous state spaces. Deep Q-Networks (DQNs) \cite{DeepQNet} overcome this limitation by employing neural networks (Q-networks) to approximate Q-values, thereby mitigating the curse of dimensionality.

Quantum Reinforcement Learning (QRL) extends these ideas by exploiting the expressive capacity of variational quantum circuits. Skolik et al. \cite{Skolik2022quantumagentsingym} proposed substituting classical neural networks with parameterized quantum circuits to enhance representational power. QRL architectures are typically hybrid in nature: quantum circuits are employed to approximate value functions or action functions, while classical processors perform gradient updates and manage environment-related computations.

\section{Methods for Framework Integration and Benchmark Development}
\label{sec:methods}

In this section, we first introduce our new architecture for the QED-C framework and demonstrate how it enables integration with other benchmarking suites and with NVIDIA CUDA-Q for high-performance simulation in Section \ref{sec:framework_integration}. Then, to highlight the extensibility made possible by these architectural advancements, we present the implementation of dynamic circuit variants in Section \ref{sec:dyn-b} and a new quantum reinforcement learning benchmark in Section \ref{sec:QRL}.

\subsection{Integration of QED-C Benchmarking Framework}
\label{sec:framework_integration}

 The QED-C suite offers numerous application-oriented benchmarks, such as Hamiltonian Simulation, Quantum Fourier Transform, and the Quantum Approximate Optimization Algorithm, with ongoing efforts to expand this set. Each benchmark is organized into subdirectories corresponding to different quantum computing APIs, including Qiskit, CUDA-Q, Cirq, and Braket, with provider-specific code used for execution. The framework allows users to run sweeps of circuits to produce benchmarking results across various circuit sizes.

While the QED-C framework works effectively as a standalone tool, the variety of existing benchmarking frameworks with different purposes presents an opportunity to enhance its impact by enabling integration with external efforts. Allowing QED-C to interoperate with these frameworks would let them benefit from the suite’s existing collection of quantum applications. However, the current QED-C framework design introduces several barriers to such integration. 
The framework currently implements a tightly coupled end-to-end execution model that processes circuit setup, execution, and analysis as a single unified pipeline, making it difficult for external frameworks to access individual components independently. Additionally, functionality such as circuit plotting, metric computation (circuit depth and width), and benchmark circuit retrieval is encapsulated within an internal interface, limiting external access to these important capabilities. Finally, the framework relies on file path-based internal imports that require specific directory structures, creating dependencies that complicate integration as a submodule.

To address these challenges, we implemented three key architectural changes to the suite:
\begin{itemize}
    \item \textit{Modularization} of the benchmark workflow
    \item A \textit{get\_circuits} flag for circuit retrieval
    \item \textit{Quantum kernels} for flexible circuit loading
\end{itemize}

First, the updated design separates the benchmarking workflow into distinct and independent stages—problem generation, execution, and analysis—that can be accessed and run independently, as shown in Figure~\ref{fig: QEDC_Architecture}. This modular framework allows external frameworks to access and run individual components as needed rather than being forced to use the complete pipeline. Second, we provide a \textit{get\_circuits} flag that enables users to retrieve benchmark circuits and their metadata without triggering execution. Third, we replaced the fragile file path-based dependencies with a more robust quantum ``kernel" architecture that dynamically loads both shared components and benchmark-API-specific components at runtime. This kernel-based architecture eliminates code duplication and enables support for additional quantum computing APIs with minimal modifications to existing benchmarks. Together, these architectural changes make the QED-C framework a flexible, maintainable package that can readily integrate with diverse external benchmarking systems.

In the following, we demonstrate these improvements by integrating QED-C with the popular benchmarking suite, \emph{pyGSTi}, and the HPC-enabling NVIDIA CUDA-Q.

\subsubsection{pyGSTi integration}

\begin{figure}[t!]
\centering
\includegraphics[width=\linewidth]{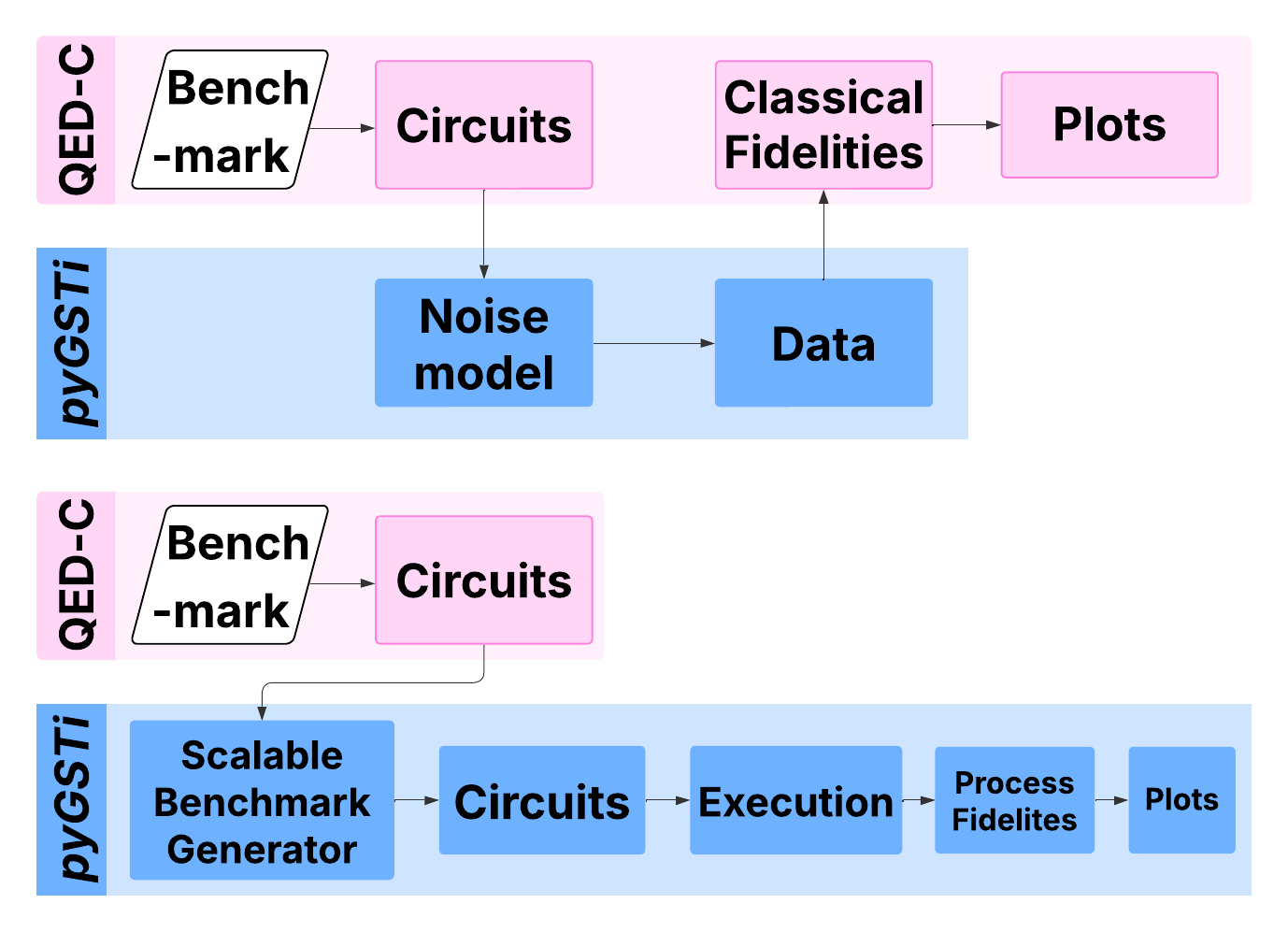}
\caption{\textbf{\textit{QED-C and pyGSTi} Integration.} Two benchmarking process flows through QED-C's suite and Sandia QPL's \textit{pyGSTi} framework. The upper workflow shows the simulation of a benchmark from the QED-C's suite using \emph{pyGSTi}'s sophisticated noise models. In this case, pyGSTi simply replaces the execution of the QED-C's circuit on real hardware or a provider's built-in simulators. The upper workflow shows using \emph{pyGSTi} to convert a QED-C benchmark into a scalable benchmark for measuring process fidelity. In this case, \emph{pyGSTi} converts the circuits created by the QED-C's suite into other circuits (using, e.g., mirror circuit fidelity estimation \cite{proctor2022establishing}) that are then executed (on real hardware or a simulator). \emph{pyGSTi} then processes the data to produce estimates of circuit process fidelities, and can create a variety of performance summary plots.} 
\label{fig: QEDC_Sandia_Integration}
\end{figure}

One of the capabilities of \textit{pyGSTi} is the ability to execute quantum circuits through customized noise models such as crosstalk errors, coherent errors, or time-varying errors. This flexibility allows users to simulate realistic hardware imperfections and study their impact on circuit execution. The QED-C benchmarking suite is leveraged to generate problem-specific quantum circuits, circuits are simulated using \textit{pyGSTi}, and then resulting fidelities are computed. The integration enables users to explore algorithm performance and error characteristics within a controlled, reproducible simulation. The steps to realize this process are illustrated in ~\autoref{fig: QEDC_Sandia_Integration}. The selected benchmark circuits are generated via the QED-C suite, processed and executed through \textit{pyGSTi}, analyzed with QED-C's fidelity calculations, and finally visualized using methods in \textit{pyGSTi}.


\subsubsection{CUDA-Q integration}

Building on the prototype CUDA-Q integration described above, we redesigned the circuit and execution implementation to make use of the new modular QED-C architecture to provide seamless CUDA-Q support. This new implementation introduces a kernel-based approach for CUDA-Q integration and enables distributed quantum simulation through MPI support for high-performance computing environments.

To eliminate the code duplication inherent in our prototype approach, we restructured each benchmark to dynamically import API-specific code (e.g., Qiskit, CUDA-Q, or Cirq) based on the selected execution option. This approach replaces the need to clone and maintain separate benchmark instances for each API. The challenges encountered during CUDA-Q integration drove the development of this new architecture, which has since been fully applied to other supported APIs, particularly Qiskit.

The API-specific code consists of two components: a generic execution module shared across all benchmarks, and a benchmark-specific quantum ``kernel". At runtime, this kernel-based architecture dynamically loads both the shared execution component and the benchmark-specific kernel for each execution. This design reduces code duplication and enables future API additions with minimal modifications to existing benchmarks.

Our MPI integration enables parallel execution of quantum kernels across multiple GPUs, with automatic state reconstruction from distributed computations. Each benchmark runs multiple instances across separate GPUs, where the rank 0 process handles initialization, logging, and analysis, while other ranks process quantum state subsets. This parallel execution is transparent to benchmark users, requiring no changes to their workflow.


\subsection{Dynamic Circuit Benchmark}
\label{sec:dyn-b}
With recent advances in dynamic circuits, we integrate two dynamic circuit benchmarks: Dynamic Quantum Fourier Transformation and Dynamic Quantum Phase Estimation into the QED-C benchmarking framework. We also propose new fidelity measurement protocols for both of these algorithms. 

\subsubsection{Dynamic Quantum Fourier Transformation}

Quantum Fourier Transform (QFT) is the quantum analogue of the discrete Fourier transform. Just as the discrete Fourier transform maps a signal from the time domain to the frequency domain, QFT transforms computational basis states into superpositions where the amplitudes encode relative phases that effectively represent the input in the quantum frequency domain.

QFT is a unitary operation on an $n$-qubit system represented as
\[
\mathrm{QFT}\,\lvert x\rangle
= \frac{1}{\sqrt{2^n}}
  \sum_{k=0}^{2^n-1}
    e^{2\pi i\,xk/2^n}\,\lvert k\rangle,
\]
which maps each computational basis state \(\lvert x\rangle\) to an equal-amplitude superposition, where each component $\lvert k\rangle$ acquires a phase factor $e^{2\pi i\,xk/2^n}$ that encodes the input $x$. QFT can be implemented in a quantum circuit with \(O(n^{2})\) elementary gates by iteratively applying a Hadamard gate to qubit \(q_{n-1-j}\)  
\(\bigl(\text{for } j = 0, 1, \dots, n-1\bigr)\),  
followed by a cascade of controlled-\(R_z\!\bigl(\pi/2^{k}\bigr)\) rotations \(\bigl(\text{for } k = 0,1, \dots, n-1-j\bigr)\), with $q_{n-1-j}$ as the control and $q_{n-1-j-k}$ as the target qubit \cite{sahin2020qftarith,ruizperez2017qftarith}.  
Optionally, one may perform a bit-reversal swap of the qubit order at the end, depending on the chosen traversal convention.  
This decomposition turns the full \(2^{n}\!\times\!2^{n}\) matrix into a circuit of depth \(O(n)\), making the QFT a cornerstone subroutine in algorithms such as phase estimation and Shor’s algorithm \cite{weinstein1999qft,Nielsen2000Quantum}. 

\begin{figure}[t!]
  \centering
  \begin{subfigure}[t]{0.48\textwidth}
    \centering
    \includegraphics[width=\linewidth]{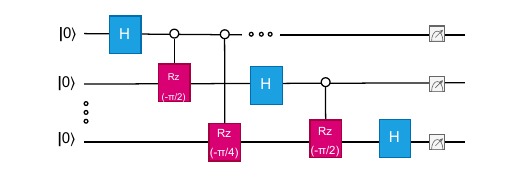}
    \caption{Static IQFT}
    \label{fig:static_iqft}
  \end{subfigure}%
  \hfill%
  \begin{subfigure}[t]{0.48\textwidth}
    \centering
    \includegraphics[width=\linewidth]{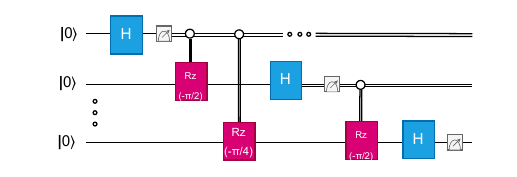}
    \caption{Dynamic IQFT}
    \label{fig:dynamic_iqft}
  \end{subfigure}

  \caption{%
    \textbf{Static vs.\ Dynamic IQFT Circuits.}
    (a) In the static circuit, the Hadamard and controlled-$R_z$ gates are applied in a fixed sequence with final measurement at the end.
    (b) In the dynamic variant, each qubit is measured immediately after the Hadamard, and the measurement result determines whether to apply the subsequent $R_z$ rotations. 
  }
  \label{fig:iqft_comparison}
\end{figure}

Inverse Quantum Fourier Transformation (IQFT) is the inverse of QFT unitary operation on an $n$-qubit system represented as \[
\mathrm{IQFT}\,\lvert k\rangle
= \frac{1}{\sqrt{2^n}}
  \sum_{x=0}^{2^n-1}
    e^{-2\pi i\,xk/2^n}\,\lvert x\rangle,
\] which maps the amplitude superposition carrying phases to computational basis states $|x\rangle$.  IQFT can be implemented similarly to  QFT, with a scalable and repeatable circuit structure, as shown in~\autoref{fig:iqft_comparison}(a). 
If IQFT is applied after QFT within the same circuit, the resulting state will be the original initial state. 

To transform the QFT or IQFT into its dynamic circuit version, the key idea is the principle of deferred measurement.  Specifically, if a measurement follows immediately after a controlled rotation, with the control qubit being measured, the controlled rotation can be deferred after the measurement as a classically controlled gate. Thus, in QFT and IQFT, all controlled-$R_z$ gates can be moved after the measurements, resulting in a dynamic circuit structure as shown in~\autoref{fig:iqft_comparison}(b). The dynamic QFT and IQFT replace all two-qubit gates with classically controlled single-qubit gates, thereby eliminating connectivity constraints entirely.

 \subsubsection{Dynamic Quantum Phase Estimation}\label{sec:dyn-qpe}
Quantum Phase Estimation (QPE) solves for the phase \(\phi\) in the eigenvalue equation $U\ket{\psi} = e^{2\pi i \phi}\ket{\psi}$ by coherently mapping \(\phi\) onto a t-qubit ancilla. It starts with the preparation of the quantum state $\frac{1}{\sqrt{2^t}}\sum_{k=0}^{2^t-1}\ket{k}\,\ket{\psi}$ with the application of Hadamard gates, and then the application of the controlled-\(U^{2^j}\) gates (each ancilla qubit \(j\) controls \(U^{2^j}\) on the eigenstate register) to imprint phases \(e^{2\pi i k \phi}\). This produces a quantum state $\frac{1}{\sqrt{2^t}} \sum_{k=0}^{2^t-1} e^{2\pi i k \phi} \ket{k}\,\ket{\psi}$. Finally, a dynamic inverse Quantum Fourier Transform is applied to the ancilla register, and measurement of this register then yields the phase $\theta$ \cite{Cleve1998Quantum}.


\subsubsection{Fidelity evaluation method}

Fidelity evaluation methods are often application dependent. Here, we propose two methods for evaluating the fidelity of QFT circuits. In the first method, we begin by converting the secret integer $s$ into its n-bit binary representation. We then initialize all the n qubits in the $|0\rangle$ state and, for each bit in the string that equals 1, we apply an X gate to flip it to the state $|1\rangle$. In this way, we prepare the quantum registers in the computational basis state $|s\rangle$. A standard QFT circuit is then applied, followed by a sequence of controlled phase rotations \(R_z(\pi/2^k)\) to prevent higher-level optimizers from canceling adjacent transforms. After this, we consider two variants: (1) applying a standard IQFT followed by final measurements, or (2) applying a dynamic IQFT, where the mid-circuit measurement results stored in the classical registers are post-processed to get the outcome distribution. The choice between these variants is controlled by a Boolean flag that selects either the static or dynamic approach. The expected output is a bit-string corresponding to $s+1$ (mod $2^n$), due to the controlled phase rotations inserted between QFT and IQFT. 

In the second method, all qubits are first prepared in an equal superposition state by applying Hadamard gates to each qubit. The secret integer $s$ is then encoded directly by applying single‐qubit phase rotations \(R_z(s\,\pi/2^k)\) proportional to each binary digit. After phase encoding, the inverse QFT is performed, again with two possible implementations: either the static IQFT with measurements or the dynamic IQFT, chosen according to the Boolean flag. The resulting measurement counts are then analyzed to infer the output, which ideally corresponds to the encoded secret integer $s$.

The benchmarking method for QPE is straightforward. We first prepare the ancilla qubits in an equal superposition and then apply a sequence of controlled operations. Finally, either a static IQFT with final measurements or a dynamic IQFT method is applied, chosen via the Boolean flag. The expected outcome should be a bitstring corresponding to the closest binary approximation of the input phase $\theta$, scaled by $2^t$, where $t$ is the number of ancilla qubits.

The measurement counts from the chosen hardware or simulator for all these methods are compared to the expected distribution to calculate the Hellinger and polarization fidelities. We also report additional metrics, including average circuit‑creation time, elapsed and execution times, average normalized transpiled depth, and circuit algorithm depth. 


\subsection{Quantum Reinforcement Learning Benchmark}
~\label{sec:QRL}
Quantum Machine Learning (QML) and Quantum High-Performance Computing (Quantum HPC) are rapidly emerging as leading directions in quantum computing. Within QML, Quantum Reinforcement Learning (QRL) has become a key subfield. QRL employs hybrid quantum–classical algorithms that interact directly with an external environment. This environment, typically implemented as a classical software module or hardware device, provides observations and executes actions specified by the quantum agent. The inclusion of this separate classical entity reshapes how researchers design quantum queues and manage quantum–classical coupling. These challenges make QRL a unique benchmark, particularly for the software stack of a quantum service provider tasked with handling quantum–classical interactions.

\subsubsection{QRL Benchmark Design}

\begin{figure}
\centering
\includegraphics[width=\linewidth]{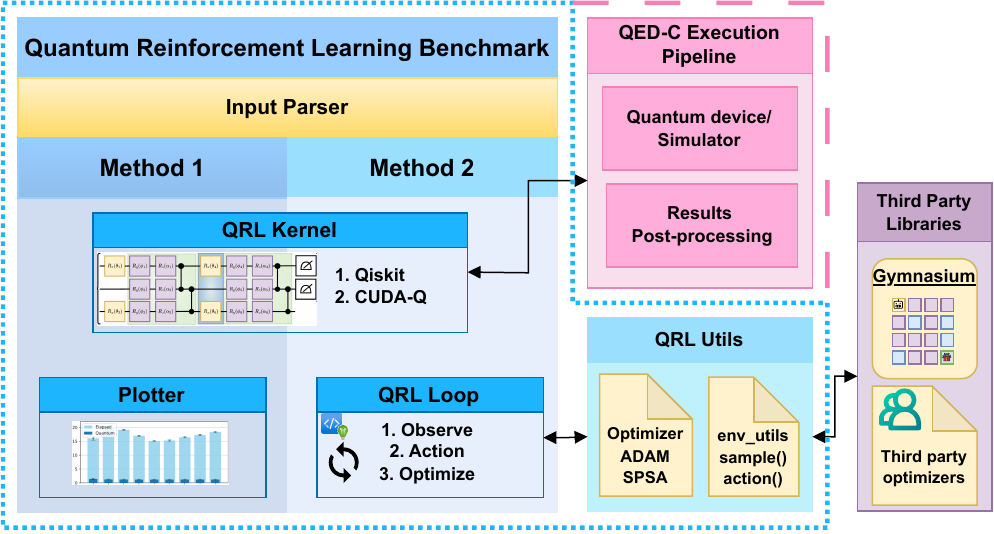}
\caption{\textbf{QRL Benchmark Overview.} Figure illustrates the main building blocks of the QRL benchmark. The components enclosed within the blue dotted line belong to the QRL benchmark, while the blocks outlined by both the pink dashed line and the blue dotted line are part of the QED-C application-oriented benchmark. The remaining elements represent third-party libraries that can be interfaced through QRL utils interface.}
\label{fig:QRL_overview}
\end{figure}

To address these challenges, we designed the QRL benchmark with two main components (see~\autoref{fig:QRL_overview}), following the overall architecture of the QC-App-Oriented Benchmark described in~\autoref{subsec:Arch}. The goal is to provide users with the flexibility of benchmarking either individual circuits or the full QRL loop, as illustrated in~\autoref{fig:QRL_overview}.  

\textbf{QRL Benchmark Program:}  
With modularity as a guiding principle, the benchmark performs four core functions:  
(1) parsing user inputs,  
(2) interfacing with the environment program,  
(3) generating and executing quantum circuits, and  
(4) optimizing ansatz parameters.  

Each function is implemented in a modular fashion to facilitate future extensions. For instance, the \texttt{env\_utils} module serves as a bridge between the benchmark and the environment program, enabling straightforward integration of additional environments. Similarly, the circuit generation, execution backends (currently supporting IBM Quantum and CUDA-Q), and optimizer modules follow the same modular structure.

To demonstrate the benchmark, we select the FrozenLake environment and apply the Quantum DQN approach. We chose these due to the simplicity of the environment and the resulting quantum circuit.

\textbf{Environment:}  
We use the FrozenLake environment from the Gymnasium library~\cite{gymnasium}. FrozenLake is a grid-based navigation task (4$\times$4 or 8$\times$8) where the agent must reach a goal tile while avoiding holes. Reaching the goal yields a reward of $+1$, while all other steps yield a reward of $0$. Despite its simplicity, FrozenLake is well suited for quantum experiments, given current qubit limitations and noise levels. A detailed description of state transitions is provided in Appendix~\ref{app:env}.

\begin{figure}
\centering
\includegraphics[width=\linewidth]{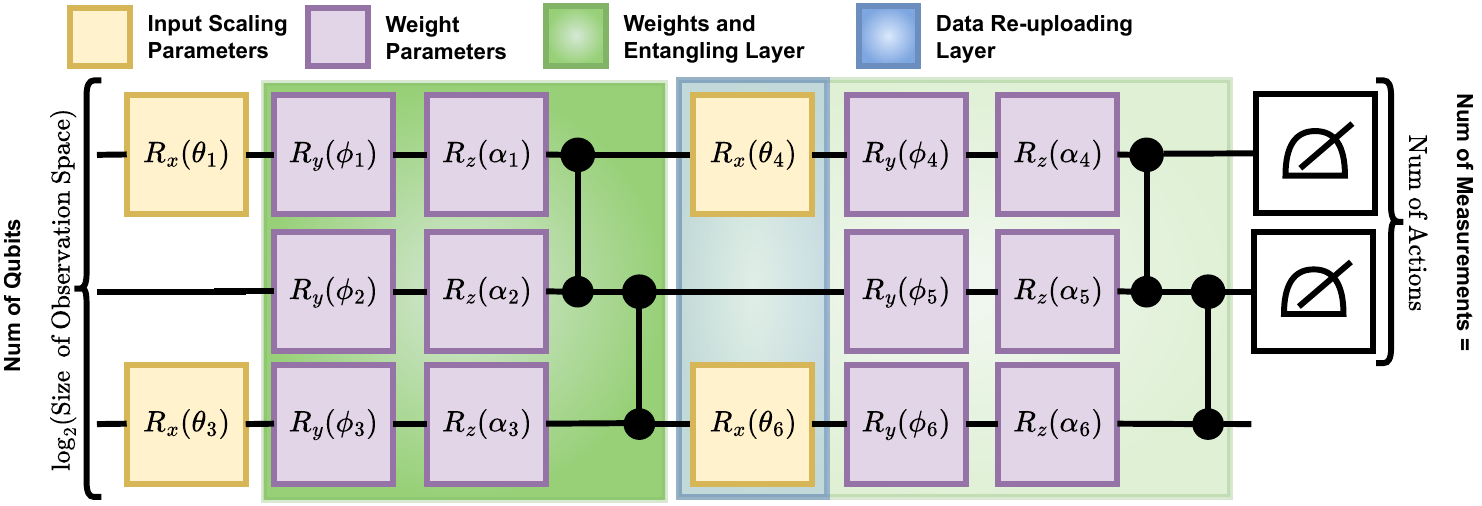}
\caption{\textbf{Ansatz Circuit Description.} Parameterized quantum circuit used in the QRL benchmark. The circuit combines $R_y$ and $R_z$ rotations with $R_x$ gates for input encoding and optional data re-uploading.}
\label{fig:PQC}
\end{figure}

\textbf{Parameterized Quantum Circuit Kernel:}  
The agent is implemented with a parameterized quantum circuit (ansatz) that acts as a function approximator for Q-values. The design follows Skolik et al.~\cite{Skolik2022quantumagentsingym} and Kruse et al.~\cite{qaio25}, and consists of three main parts, refer to ~\autoref{fig:PQC}:
\begin{enumerate}
  \item Input encoding with $R_x$ gates,
  \item Variational layers of $R_y$ and $R_z$ rotations with $CZ$ entangling gates,
  \item Optional data re-uploading layers~\cite{perez2020data}.
\end{enumerate}
Unlike algorithms such as VQE or QAOA, where all qubits are measured, here only the qubits corresponding to the number of actions are measured. For example, an 8$\times$8 FrozenLake grid requires 6 qubits to encode the state but measures only 4 qubits, one for each action. With the provision of an optional data re-uploading parameter, the QRL benchmark is now the only benchmark within the QC-App-Oriented framework that supports data re-uploading. This gives users the ability to test the performance of the data re-uploading technique, which has been argued to improve results in QML. Studies such as Barthe and Perez-Salinas \cite{Barthe2024gradientsfrequency} explore the efficacy and design of Quantum Reuploading Models (QRUs). Our benchmark enables researchers to conduct similar studies through a togglable data re-uploading parameter. Further details on the ansatz, including the encoding of inputs, are provided in Appendix~\ref{app:ansatz}.

To support both circuit-only execution and the full QRL loop, the benchmark provides two modes:
\begin{itemize}
  \item \textbf{Method 1 (Circuit benchmarking):} Generates and evaluates PQC ansatz circuits across a range of qubit sizes. 
  \item \textbf{Method 2 (Full QRL workflow):} Executes the complete DQN-QRL loop on FrozenLake. The agent follows an $\epsilon$-greedy exploration strategy, stores experiences in a replay buffer, and updates parameters at fixed intervals. Supported optimizers include SPSA~\cite{spall1992multivariate} and ADAM~\cite{kingma2014adam}. Detailed parameter settings are provided in Appendix~\ref{app:training}.
\end{itemize}

Details about user-configurable parameters for both methods are included in Appendix~\ref{app:inputs}.

\subsubsection{Model Performance Evaluation}

\textbf{Method 1:} In Method 1, the generated circuits are executed on a noiseless simulator, followed by either a noisy simulator or a user-specified quantum device. The resulting outputs report fidelities, circuit depth, runtime, and volumetric plots, consistent with the performance metrics of established benchmarks such as QFT and QPE.

\textbf{Method 2:} Evaluation tracks both learning progress and quantum–classical resource usage. During training, per-step logs record circuit counts, exploration versus exploitation, and environment interactions. At the end of training, aggregate metrics include total steps completed, gradient evaluations, episode counts, success rates, returns, and timing breakdowns, reported as both plots and console outputs. Definitions of these metrics are provided in Appendix~\ref{app:metrics}.

\section{Results and Analysis}
\label{sec:results-analysis}

\subsection{Experimental Setup}

To demonstrate our integration efforts, we present results from \textit{pyGSTi} for advanced noise simulation and CUDA-Q \cite{CUDA-Q} for multi-GPU simulations using our QFT benchmark. We evaluate the \textit{pyGSTi} integration utilizing its built-in simulation model, and the CUDA-Q integration with the Lawrence-Berkeley National Laboratories NERSC Perlmutter supercomputer \cite{Perlmutter}.

The extensibility of the new QED-C architecture is evaluated by running the dynamic circuit variations on IBM Pittsburgh, and our new QRL benchmark on IBM Torino (IBM Open Access Plan) and Qiskit simulators. Note that we run method~2 of the QRL benchmark on simulators due to its rampant scaling.  

We use various metrics from the QED-C framework to evaluate benchmark executions, including execution time, fidelity, and circuit depth. Specifically, the framework supports two methods for fidelity calculation (Hellinger and Polarization), two measures of circuit depth (algorithmic and normalized), and two types of execution time (total elapsed and on-quantum-hardware) \cite{lubinski2023_10061574
,lubinski2023optimization,lubinski2024quantumalgorithmexplorationusing,chatterjee2024comprehensivecrossmodelframeworkbenchmarking,niu2025practicalframeworkassessingperformance}.

\subsection{QED-C Integration Results}

We present the results of integration with an external benchmarking tool and a new circuit API. First, we highlight the effects of crosstalk errors on QFT with \textit{pyGSTi}, then we showcase the integration results with CUDA-Q for HPC simulations.

\subsubsection{QED-C and pyGSTi}

\begin{figure}[t!]
  \centering
  \includegraphics[width=\linewidth]{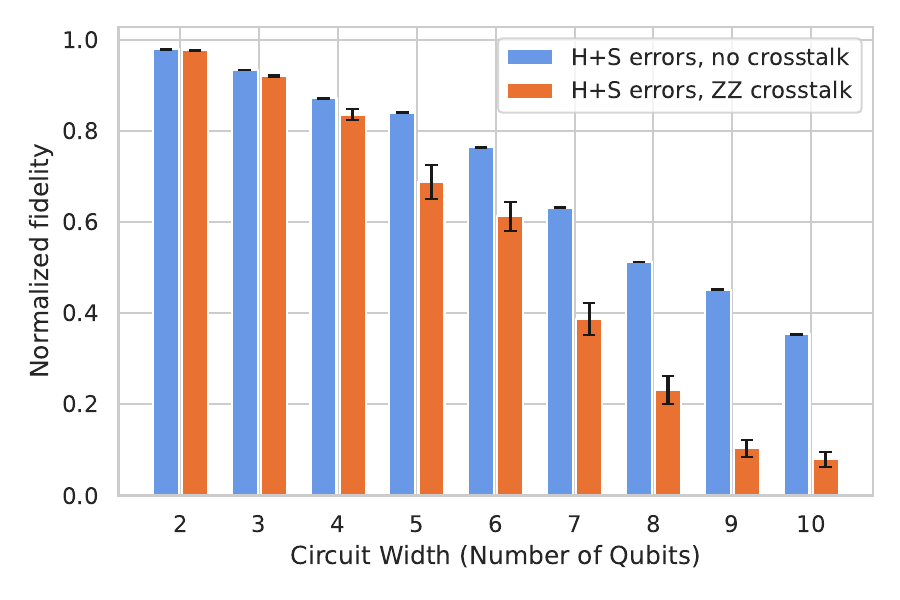}
  \caption{\textbf{Noise Analysis using PyGSTi.} Average fidelity for QFT method~1 on 2--10 qubits under a crosstalk-free noise model and $ZZ$-crosstalk noise model. Error bars are 1 standard deviation calculated from a non-parametric bootstrapped distribution.}
  \label{fig:pygsti-qft-fidelity}
\end{figure}

Using \emph{pyGSTi}'s noise modeling capabilities, we simulated the circuits of the QFT benchmark (method 1) in the QED-C benchmarking suite under a complex, realistic noise model. To demonstrate the importance of studying realistic noise, we considered two noise models: one with crosstalk and one without. Crosstalk is an important source of error in many contemporary quantum computing systems \cite{Proctor2025-cd, proctor2022measuring}, but is not included in many simple noise simulators. Our error models included local coherent and stochastic errors (see Appendix~\ref{app:noisemodels} for details) and, for the model with crosstalk, we also included coherent $ZZ$ couplings between a qubit and all its neighbors whenever a two-qubit gate is applied to that qubit. We used this example as unwanted $ZZ$ couplings are common in transmon qubits.

The processor we simulated consisted of qubits in a $12 \times 12$ grid topology with a $\{X, SX, RZ, CX\}$ gate set. This gate set and processor size are similar to contemporary quantum hardware. We transpiled the QFT benchmarking circuits to this processor and simulate their execution under each noise model. We did this for benchmarking circuits up to $n=10$ qubits and with a variety of parameters for the benchmarking circuit (in this case, $m$ different input integers at each circuit width, where $m = \min(2^n, 10)$, except that $m=8$ for $n=4$). The results of the QFT benchmark for each error model are shown in~\autoref{fig:pygsti-qft-fidelity}. 
We observe that the coherent $ZZ$ crosstalk error significantly reduces the average fidelity for all $n$, and it also increases the variance in the fidelity (over different QFT instances).

\vspace{0.3cm}

\subsubsection{QED-C and CUDA-Q Multi-GPU Quantum Simulation}

\begin{figure}[!t]
\includegraphics[width=0.96\columnwidth]{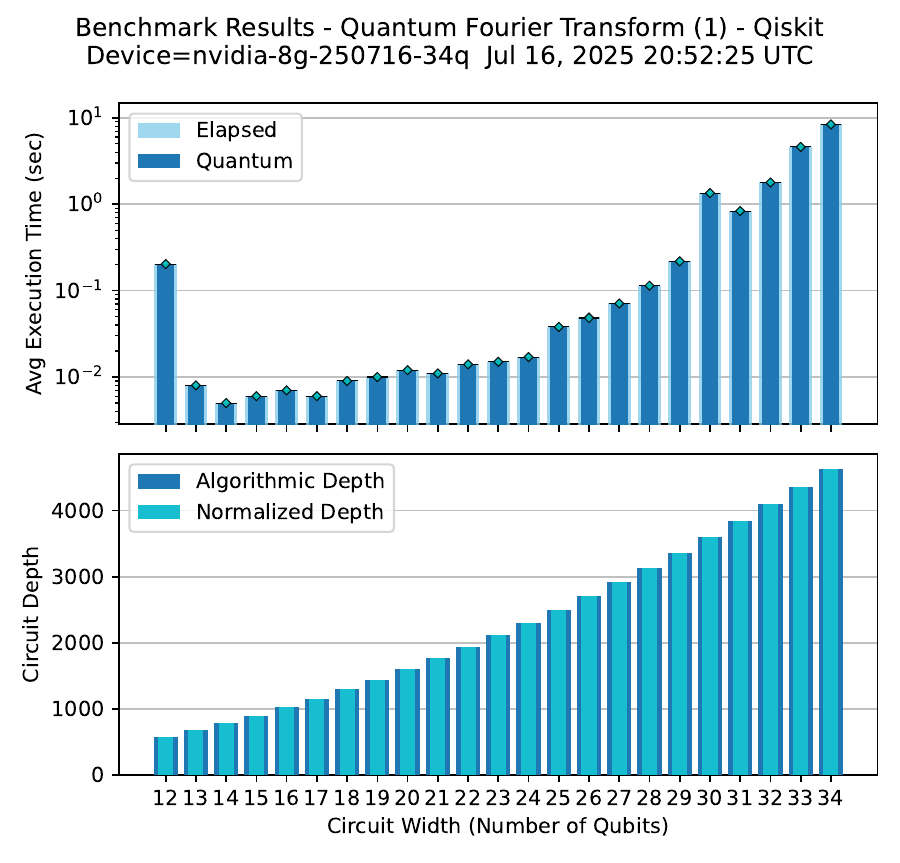}

\caption{\textbf{QED-C Benchmark Execution on multiple GPUs using CUDA-Q and MPI.} Execution time and depth profiles for the QFT algorithm at increasing problem size, from 12 to 34 qubits, using 8 Perlmutter A100 GPUs. Scaling to this number of qubits is enabled via the Message Passing Interface (MPI), enabling multi-processing and distributed memory across the GPUs. Execution time grows by a factor of 2 for each qubit increase in the problem size, a characteristic of classically implemented quantum simulation. \emph(Data collected on NERSC Perlmutter GPU partition)}
\label{fig:perlmutter-1}
\end{figure}

Classically implemented simulation of quantum circuits on NVIDIA GPUs is critically important within the context of HPC. As part of the work described in this manuscript, we enhanced the CUDA-Q execution pipeline to enable quantum simulations that span the memory and processing space over multiple GPUs using the MPI~\cite{MPI}, which enables multi-processing in distributed memory systems.

To demonstrate this integration and the power of multi-processing, we present in~\autoref{fig:perlmutter-1} the execution of our QFT benchmark on the Lawrence-Berkeley National Laboratories NERSC Perlmutter supercomputer~\cite{Perlmutter}. This benchmark was executed over qubit widths ranging from 12 to 34 qubits. Note that the execution time grows by a factor of 2 for each qubit increase in the problem size. This is a consequence of the increase in the dimensionality of the $N \times N$ matrix required for classical implementation of quantum simulation. The increase in circuit depth illustrated in the figure has minimal impact on execution time compared to the qubit scaling, as the CUDA-Q simulator applies gate operations sequentially to the quantum state vector rather than constructing large circuit matrices.

Scaling to this number of qubits (34) was made possible via the MPI coordination of memory and processing space of the CUDA-Q simulation of the QFT circuit executed in parallel on 8 A100 GPUs (on 2 nodes). Within the QED-C benchmark code, it was only necessary to enable the MPI API at the start and end of the benchmark execution and the CUDA-Q NVIDIA driver used MPI to share memory between the 8 GPUs.

\subsection{Dynamic Circuit Results}

We execute dynamic circuits on the IBM Pittsburgh hardware in sweeps of increasing circuit width, ranging from 3 to 10 qubits for the QFT and QPE benchmarks, with three similar circuits submitted per width. For each width, the batch of circuits shares the same logical structure, with only the rotation angles differing according to the randomly generated secret integer. Each circuit is executed 1000 times, and the Hellinger fidelity is used to evaluate performance. The reported average fidelities, along with other metrics such as average execution time, are computed by averaging over all circuits of the same width within each configuration.

\begin{figure}[t!]
  \centering
  \includegraphics[width=0.89\columnwidth]{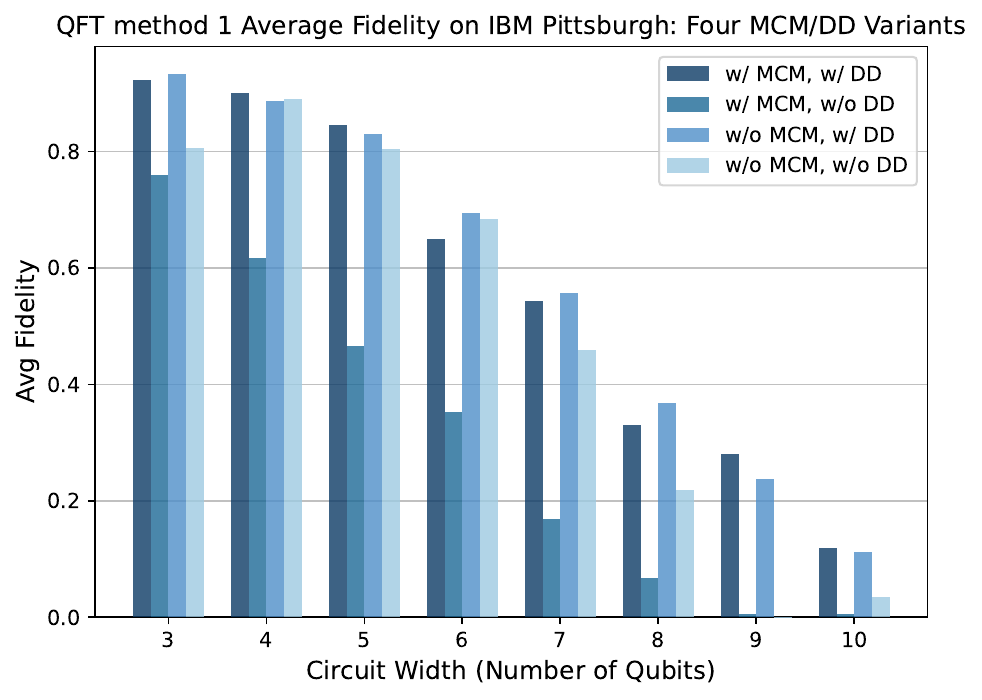}
  \caption{\textbf{QFT Method 1 Fidelity Comparison.} Average fidelity for 3--10 qubits on IBM Pittsburgh, comparing configurations with/without MCM and with/without DD. \emph(Data collected using IBM Quantum Cloud)}
  \label{fig:qft-fidelity}
\end{figure}

The higher errors inherent with MCM and the longer durations of MCM and feed-forward operations introduce additional decoherence during execution. To counter this, we apply dynamical decoupling (DD) to the circuits to mitigate these effects. DD is an error suppression technique aimed at reducing decoherence errors ~\cite{souza2012robust,niu2022effects,khodjasteh2005fault}.

We consider four experimental configurations: (i) with mid-circuit measurements and dynamical decoupling (w/ MCM, w/ DD), (ii) with mid-circuit measurements and without dynamical decoupling (w/ MCM, w/o DD), (iii) without mid-circuit measurements and with dynamical decoupling (w/o MCM, w/ DD), and (iv) without mid-circuit measurements and without dynamical decoupling (w/o MCM, w/o DD). 

\begin{figure}[t!]
  \centering
  \includegraphics[width=0.85\columnwidth]{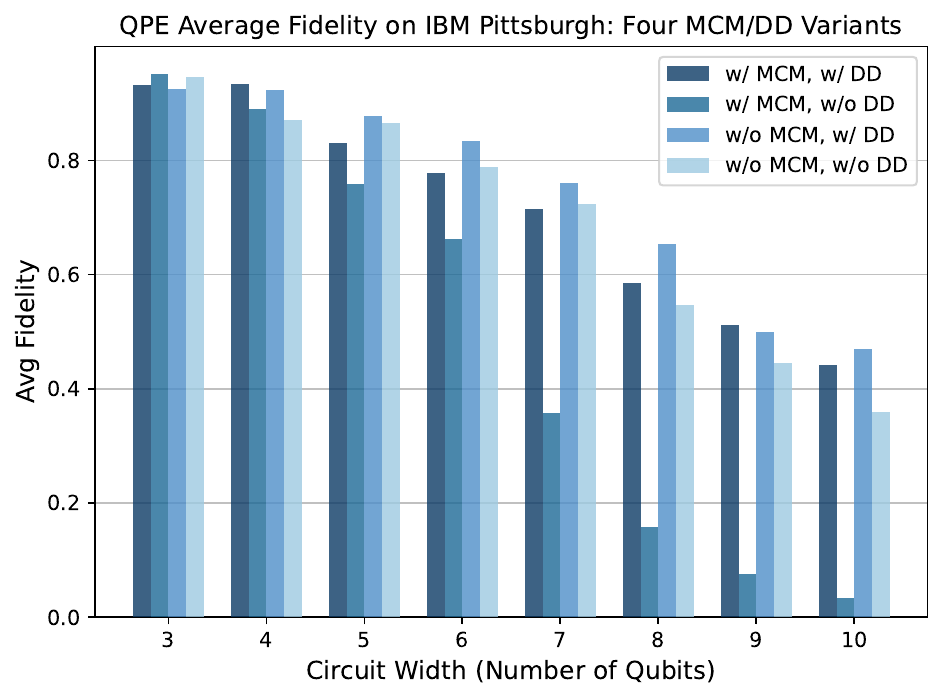}
  \caption{\textbf{QPE Fidelity Comparison.} Average fidelity for 3--10 qubits on IBM Pittsburgh, comparing configurations with/without MCM and with/without DD. \emph(Data collected using IBM Quantum Cloud)}
  \label{fig:qpe-fidelity}
\end{figure}

The results in Figures~\ref{fig:qft-fidelity} and~\ref{fig:qpe-fidelity} present the average fidelities for the QFT method~1 and QPE benchmarks executed on the IBM Pittsburgh (ibm\_pittsburgh) quantum device. For smaller circuit widths ($\sim$3 qubits), all four configurations achieve relatively high fidelities. As the circuit width increases, performance differences become more noticeable, with the w/ DD configurations consistently yielding higher fidelities in both QFT method~1 and QPE. The w/ MCM, w/ DD variant outperforms the w/o MCM, w/ DD variant at certain widths—for example, at widths 4, 5, and 9 in QFT method~1, and at widths 4 and 9 in QPE. Such higher fidelity is not uniform across all widths, indicating that the performance is hardware-dependent and shaped by factors such as readout error rates, two-qubit gate error rates, and variability in the error rates of physical qubits. Although mid-circuit measurements are often thought to introduce measurement delays that could reduce fidelity due to measurement error rates, our results indicate that when combined with dynamical decoupling, circuits with MCM can perform on par with their static counterparts, and sometimes even better. Moreover, these results underscore the value of dynamical decoupling, with or without mid-circuit measurements, in sustaining higher fidelities at larger circuit widths. An additional result on the average execution time of these quantum circuits is provided in Appendix~\ref{appx:dynamic-circuits-exec-time}.

\subsection{Quantum Reinforcement Learning Results}
We present results for both Method 1 and Method 2 of the QRL benchmark. Method 1 involves the standalone evaluation of quantum circuits employed within the QRL application, whereas Method 2 comprises the execution of the complete QRL benchmark on the 4$\times$4 FrozenLake environment. Method 1 experiments were carried out on IBM Torino, while Method 2 was performed on noisy simulator using the default noise model in the QED-C Application oriented benchmark. The empirical observations from Method 1 provide a hardware-grounded baseline that contextualizes the simulator-based outcomes of Method 2 and motivates the conclusions drawn from these experiments.

\subsubsection{The QRL ansatz benchmarking (Method 1)}
In method 1 we generated three instances of ansatz for each width of ansatz (number of qubits) with the number of layers fixed to 5. The reasoning behind the number of layers set to 5 is that this is the minimum number of layers we needed in method 2 to observe the QRL ansatz start learning reliably. 

\begin{figure}[t!]
    \centering
    \includegraphics[width=0.9\columnwidth]{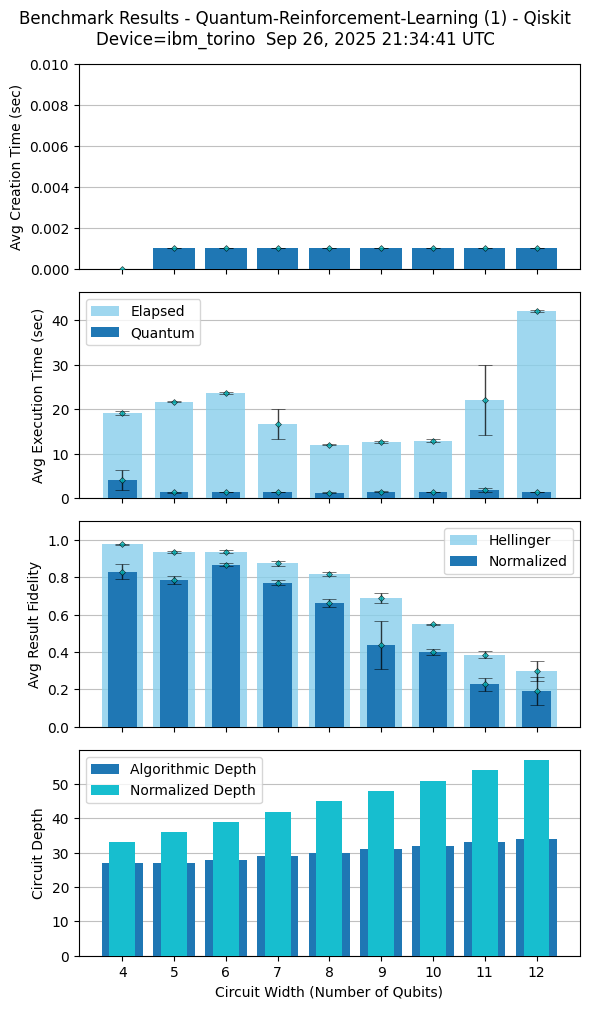}
    \caption{\textbf{QRL Method 1 Benchmarking.} Average fidelities, execution times, and circuit depths obtained for the QRL ansatz with qubit counts ranging from 4 to 12. The experiments were carried out on the \texttt{ibm\_torino} quantum processor, where for each circuit width three independent circuits were executed to ensure statistical reliability. \emph(Data collected using IBM Quantum Cloud Open Access Plan)}
    \label{fig:qrl_ansatz_fidelity}
\end{figure}
 ~\autoref{fig:qrl_ansatz_fidelity} illustrates a linear decrease in both the Hellinger distance and normalized fidelity as the qubit count increases. This behavior arises because, while the number of ansatz layers remains fixed, the circuit depth scales linearly with the number of qubits. Notably, the execution time remains approximately constant, indicating that the increase in circuit size does not translate into a measurable increase in quantum execution time.

\subsubsection{The QRL application benchmarking (Method 2)}
In Method 2, we evaluate the complete QRL loop using the FrozenLake environment provided by the gymnasium library. The experimental configuration employs a parameterized quantum circuit with three layers in the ansatz. We select this depth to maintain a balance between expressibility and runtime efficiency, particularly when using the ADAM optimizer. Data re-uploading is enabled to improve the representational capacity of the quantum model across the 200-step training horizon. The training procedure follows a staged schedule: the agent performs mostly exploration steps during the initial 100 steps, after which parameter optimization is initiated. From that point onward, training updates occur at regular intervals of 10 steps. This schedule allows the optimizer to refine the policy without incurring prohibitive overhead during the early phase of the experiment. This is due to the fact that exploration - exploitation follows an $\epsilon$ greedy exploration schedule with $\epsilon = 0.5$.

\begin{figure}[!t]
    \centering
    \includegraphics[width=\columnwidth]{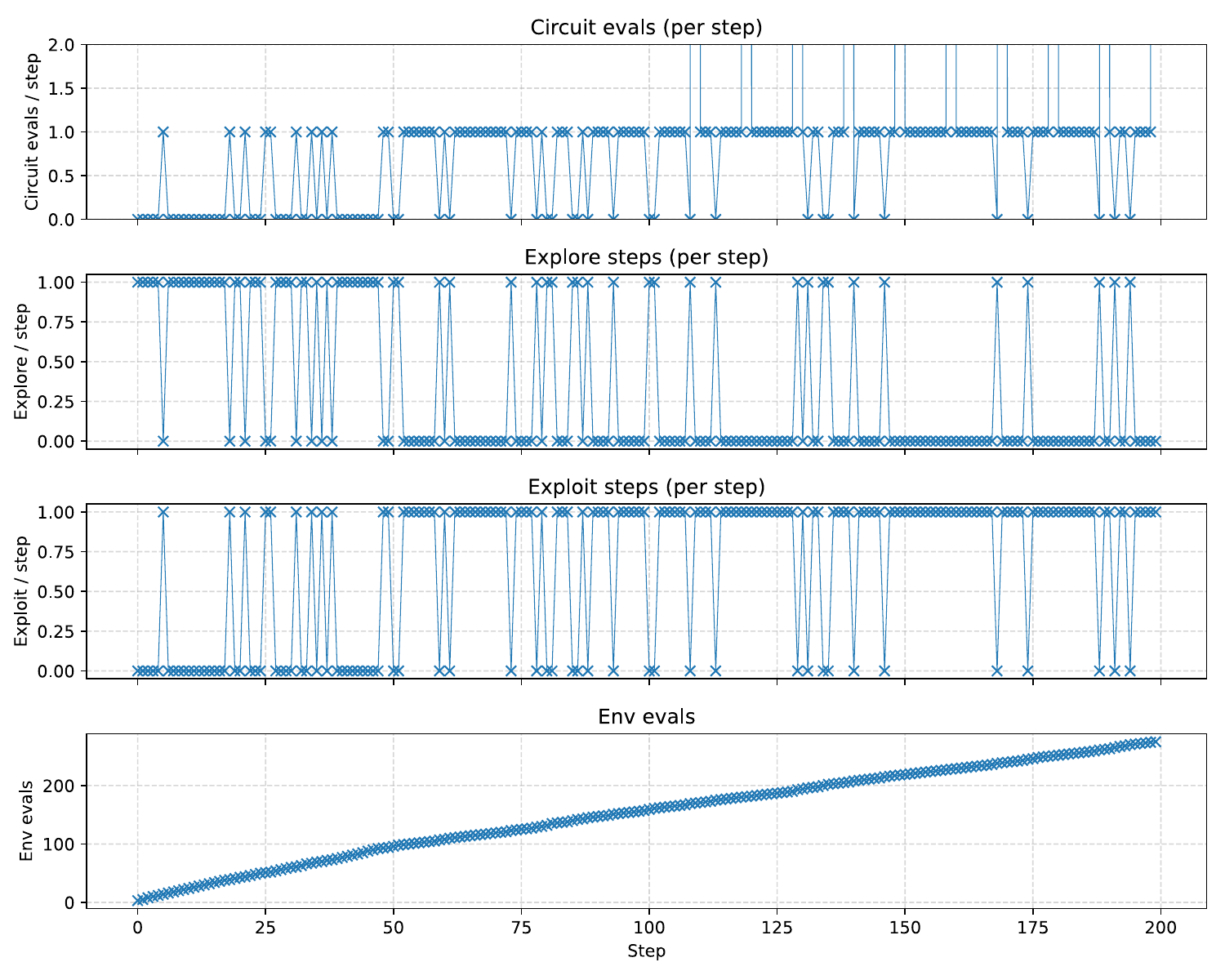}
    \caption{\textbf{QRL Method 2 Benchmark.} For each step $t$, we report circuit evaluations, exploration and exploitation steps, and the cumulative environment evaluations up to step $t$.
    \emph(Data collected using noisy simulator with default noise model of QED-C Application Oriented Benchmarks.)}
    \label{fig:QRL_timing_plot}
\end{figure}

\begin{figure}[t!]
    \centering
    \begin{subfigure}{0.49\columnwidth}
        \centering
        \includegraphics[width=0.975\columnwidth]{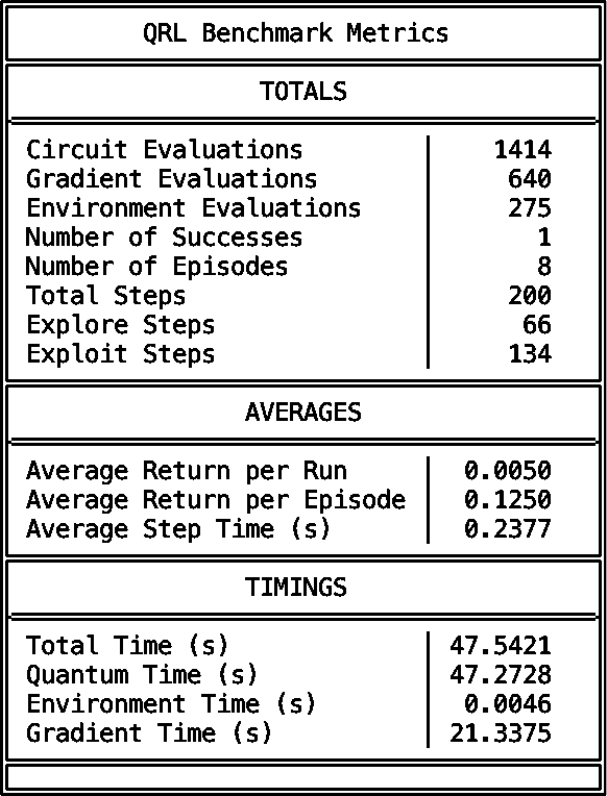}
        \caption{SPSA}
        \label{fig:a}
    \end{subfigure}
    \hfill
    \begin{subfigure}{0.49\columnwidth}
        \centering
        \includegraphics[width=\columnwidth]{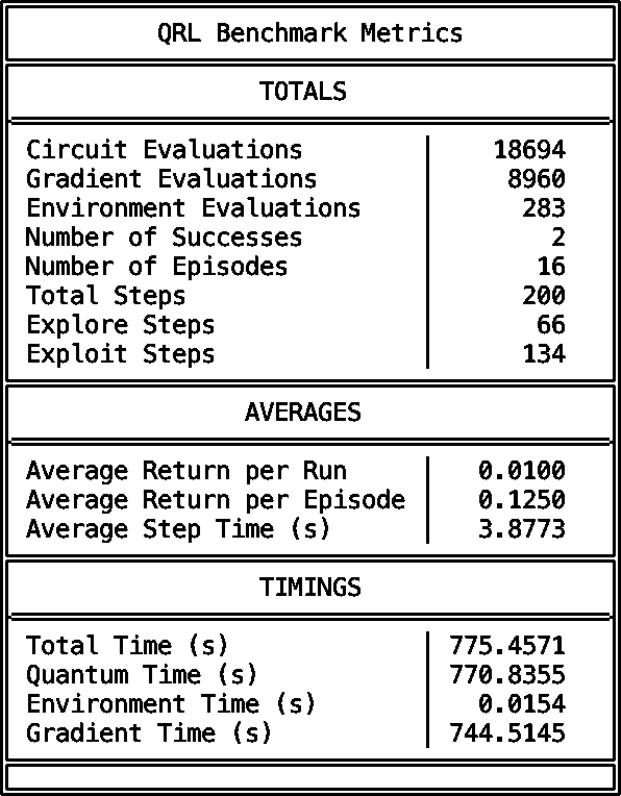}
        \caption{ADAM}
        \label{fig:b}
    \end{subfigure}
    \caption{\textbf{QRL Method 2 Benchmarking.} Final console outputs summarizing all collected statistics across 200 steps for (a) the SPSA optimizer and (b) the ADAM optimizer.
    \emph(Data collected using noisy simulator with default noise model of QED-C Application Oriented Benchmarks.)}
    \label{fig:Final_statistics}
\end{figure}

Upon completion of the benchmark runs, we record two complementary sets of outputs. First, we visualize the per-step statistics, including circuit evaluations, exploration and exploitation transitions, and environment interactions, as illustrated in~\autoref{fig:QRL_timing_plot}. Second, we report the cumulative outcomes in the final console output, which summarizes all key performance indicators collected over 200 steps. These aggregated results are presented in Figures~\ref{fig:Final_statistics}(a) and~\ref{fig:Final_statistics}(b), corresponding to the runs with SPSA and ADAM optimizers, respectively.

\autoref{fig:QRL_timing_plot} highlights the number of circuit evaluations per step, the distinction between exploration and exploitation phases, and the cumulative environment interactions. The results show that environment interactions increase nearly linearly with time, while circuit evaluations occur exclusively during exploitation steps. After step 100, sharp spikes exceeding two circuit evaluations appear, reflecting additional gradient evaluations required by the optimizer.

Due to the stochastic nature of the $\epsilon$-greedy exploration schedule, circuit evaluations in the early phase remain sparse. Consequently, executing the benchmark on reserved quantum resources results in underutilization for roughly 33\% of the runtime when $\epsilon = 0.5$, potentially incurring costs without active usage. Conversely, deferring circuit executions through standard job queues may introduce delays that are impractical for time-sensitive reinforcement learning tasks. A priority-queue scheduling strategy, akin to IBM’s \texttt{batch} mode \cite{Constantinescu_Yu_Wack_Silva_2024}, provides a more efficient alternative by ensuring both timely execution and effective resource utilization.

Figure \ref{fig:Final_statistics} presents the console output generated at the completion of QRL Method 2. The figure reports the final collected statistics for (a) SPSA and (b) ADAM optimizers, each executed for 200 steps on the noisy simulator. While the total steps, exploration steps, and exploitation steps remain identical across both optimizers, the number of circuit evaluations under ADAM is approximately 13× higher than under SPSA. This outcome is consistent with the use of the parameter-shift rule in ADAM, which requires two circuit evaluations per parameter for gradient estimation.

Based on the execution time observed for a four-qubit circuit in~\autoref{fig:qrl_ansatz_fidelity} (approximately one second per evaluation), the ADAM run requires an estimated $18{,}694$ seconds ($\sim311$ minutes) of quantum runtime, compared to only $\sim23$ minutes for SPSA. Despite this significant runtime overhead, the ADAM-trained agent reaches the goal state nearly twice as often as the SPSA-trained agent, highlighting a clear tradeoff between solution quality and execution time.

\section{Summary and Conclusion}
\label{sec:summary-and-conclusions}
 
In this work, we identify and formalize a modular interface that enables interoperability between previously incompatible benchmarking frameworks. This standardized approach reduces ecosystem fragmentation while retaining the flexibility needed for specialized optimizations across different quantum computing platforms. The modularized architecture enables researchers and practitioners to utilize selected components that depend on their specific needs, whether it is the complete integrated suite or integrating individual modules with external tools.

We validated this modular approach by integrating the QED-C Application-Oriented Benchmark suite with an established quantum benchmarking framework, \textit{pyGSTi}, for advanced circuit analysis and CUDA-Q for multi-GPU HPC simulations. The system's extensibility is demonstrated through the implementation of dynamic circuit variants and a new QRL benchmark, which are readily available across multiple execution and analysis methods within the improved QED-C suite.

The significance of this work extends beyond technical improvements to the QED-C suite. By providing standardized interfaces for application-oriented benchmarking that span algorithmic implementation to transpilation and execution, our architecture contributes to a more unified approach within this domain of quantum system benchmarking. This addresses a barrier facing the quantum computing community, where the proliferation of independent application-level benchmarking approaches has created substantial challenges for comparative analysis and framework integration.

While substantial progress has been achieved, continued development is necessary to realize the full potential of this approach. The new architecture remains under active development, and not all benchmarks have been fully migrated to support third-party integrations. Additionally, refinements to our quantum reinforcement learning benchmark and dynamic circuit implementations will enhance their practical utility. Through ongoing collaborations with benchmark partners, we are working to finalize the QED-C suite packaging and encourage community contributions to further benchmark development.

\section*{Code Availability}
\label{sec:data_and_code_availability}

The code for the benchmark suite described in this work is available at 
\href{https://github.com/SRI-International/QC-App-Oriented-Benchmarks}{https://github.com/SRI-International/QC-App-Oriented-Benchmarks}.
Detailed instructions are provided in the repository.

\section*{Acknowledgment}
The Quantum Economic Development Consortium (QED-C), comprised of industry, government, and academic institutions with NIST support, formed a Technical Advisory Committee (TAC) to assess quantum technology standards and promote economic growth through standardization. The Standards TAC developed the Application-Oriented Performance Benchmarks for Quantum Computing as an open-source initiative with contributions from multiple QED-C quantum computing members.
We thank QED-C members for their valuable input in reviewing and enhancing this work. Funding for AG and HPP was provided by the Quantum Economic Development Consortium.  Funding for NP was provided by Unitary Foundation and Quantum Computing Data.

We acknowledge the use of IBM Quantum services for this work. The views expressed are those of the authors and do not reflect the official policy or position of IBM or the IBM Quantum team.
IBM Quantum. https://quantum-computing.ibm.com, 2024.
Access to IBM systems was provided by Quantum Computing Data.

This research used resources of the National Energy Research Scientific Computing Center (NERSC), a U.S. Department of Energy Office of Science User Facility located at Lawrence Berkeley National Laboratory, operated under Contract No. DE-AC02-05CH11231 using NERSC award m4976. Computations were performed on the GPU partition of the Perlmutter supercomputer.

This material is based upon work supported by the U.S. Department of Energy, Office of Science (DE-FOA-0002253), National Quantum Information Science Research Centers, Quantum Systems Accelerator. T.P. acknowledges support from an Office of Advanced Scientific Computing Research Early Career Award. Sandia National Laboratories is a multi-program laboratory managed and operated by National Technology and Engineering Solutions of Sandia, LLC., a wholly owned subsidiary of Honeywell International, Inc., for the U.S. Department of Energy's National Nuclear Security Administration under contract DE-NA-0003525. All statements of fact, opinion, or conclusions contained herein are those of the authors and should not be construed as representing the official views or policies of the U.S. Department of Energy or the U.S. Government.

\bibliographystyle{ieeetr}
\bibliography{references_dynamic}

\clearpage
\appendix

\section{Appendix}
\label{appendix}

\begin{figure}[t!]
  \centering
  \includegraphics[width=0.8\columnwidth]{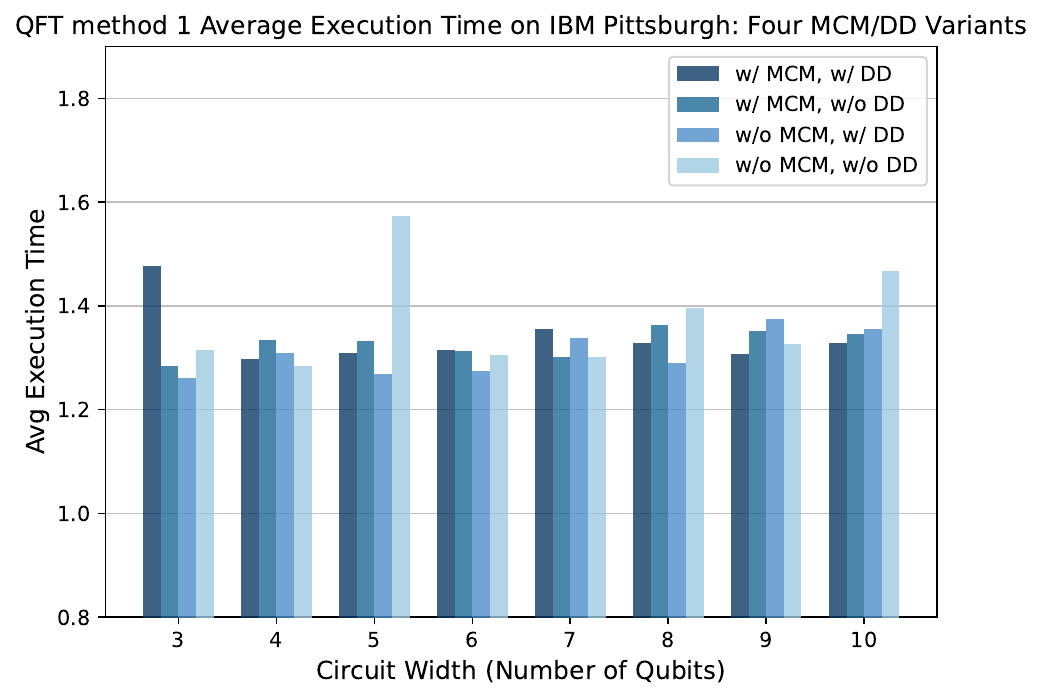}
  \caption{\textbf{QFT Method 1 Execution Times.} Average execution time for the QFT method 1 on 3–10 qubits on IBM Pittsburgh hardware, comparing configurations with/without MCM and with/without DD. \emph(Data collected using IBM Quantum Cloud)}
  \label{fig:qft-execution_time}
  
\end{figure}

\subsection{Dynamic Circuits - Average Execution time}
\label{appx:dynamic-circuits-exec-time}

The results in Figure~\ref{fig:qft-execution_time} show that the average execution time for the smallest circuit width (2 qubits) is noticeably higher for the w/ MCM, w/ DD variant, which is likely due to the job queue being submitted in the ascending order of circuit width. The initial execution likely incurs additional latency as the hardware is getting prepared for the first job. For all subsequent circuits, the execution times range from approximately 1.2-1.6 seconds across all four variants, indicating a near constant overhead. This stable range appears to represent the minimum achievable runtime for smaller and mid-sized circuits in this experimental run. Even though minor fluctuations appear between the variants, no consistent advantage or disadvantage is evident from the use of mid-circuit measurement or dynamic decoupling.

\subsection{Noise model details}\label{app:noisemodels}

Here we specify the details of the noise models used in our pyGSTi simulations of the QFT benchmark. These noise models are defined using the elementary error generator framework~\cite{Blume-Kohout2022-ln}. In this framework each gate $G$ is modeled as
\begin{equation} \label{eq:post-gate-error-gen}
    G = e^L \bar{G}.
\end{equation}
Here, $G$ is the action of the ideal gate and $L$ is an error generator for the post-gate error channel. We consider a noise model where $L$ can be expressed in terms of Hamiltonian and stochastic elementary error generators:
\begin{align}
    L &= \sum_P h_P H_P  + \sum_P s_P S_P,
\end{align}
where
\begin{align}
    H_P \left[ \rho \right] &= -i \left[ P, \rho \right], \\
    S_{P} \left[ \rho \right] &= P \rho P - \rho .
\end{align}
Here, $P$ is a $w$-qubit non-identity Pauli operator, where $w$ is the number of qubits acted on by $G$. The Hamiltonian errors $H_P$ describe coherent errors, and the stochastic errors $S_P$ describe the probabilistic application of a Pauli operator.

For the error model that we simulated, we sampled Hamiltonian error rates $h_P$ for the X, SX, and CX gates from a normal distribution with standard deviation $\sigma = 5 \cdot 10^{-4}$. The stochastic error rates $s_P$ are sampled for the $X$, $SX$, and $CX$ gates uniformly from $[0, 5 \cdot 10^{-4}]$. We do not add noise to the $RZ$ gates because they are implemented virtually in most quantum computing systems, and therefore are widely expected to be error-free. The crosstalk noise model uses the same Hamiltonian and stochastic error rates as the crosstalk-free model, but it introduces a crosstalk error on the $CX$ gate. Specifically, every qubit adjacent to the control or target qubit in the CX gate experiences an $H_{ZZ}$ error with the control or target qubit, respectively, with error rate $h_{ZZ} = 10^{-2}$.

The particular error model that we used here is an example intended to illustrate the importance of modelling crosstalk errors and the capabilities of the \emph{pyGSTi} noise simulator. \emph{pyGSTi} can easily create a wide range of error models defined in terms of the elementary error generators.

\subsection{Additional Details for the QRL Benchmark}
\label{app:qrl_details}

\subsubsection{FrozenLake Environment}
\label{app:env}

\begin{figure}[t!]
    \centering
    \includegraphics[width=0.6\columnwidth]{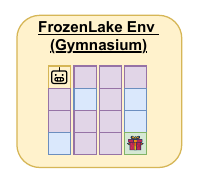}
    \caption{\textbf{FrozenLake Environment.} Example of a 4x4 FrozenLake environment provided by Gymnasium Library. Here frozen tile is purple, hole tiles are depicted in blue and green denotes the goal tile.}
    \label{fig:frozenlake}
\end{figure}

FrozenLake (Gymnasium~\cite{gymnasium}) is a simple but effective test environment for Reinforcement Learning (RL) tasks. It is a grid of size 4$\times$4 or 8$\times$8 (which we have kept user-settable). An example of a 4x4 FrozenLake environment in shown in ~\autoref{fig:frozenlake}. Each tile is either:
\begin{itemize}
  \item a \textbf{frozen tile} (safe, episode continues),
  \item a \textbf{hole} (episode ends),
  \item the \textbf{goal} (reward $+1$, episode ends).
\end{itemize}
The agent starts at the top-left tile. At each step, it can move up, down, left, or right. Falling into a hole or leaving the grid ends the episode with a reward $0$. Safe tiles also yield $0$ but allow the episode to continue. The goal tile gives a reward of $+1$ and ends the episode and sets a \textit{done} flag, which in turn increases the success counter. Though small, and given the goal is to benchmark the hardware capabilities, this environment is well-matched to current hardware limits and can be further scaled as the api's in \texttt{env\_utils} have been modularized to support additional randomized maps.

\subsubsection{Inputs and Parameter Summary}
\label{app:inputs}

For completeness, the user-settable benchmark parameters are summarized in the tables  ~\autoref{tab:cli_m1}  and ~\autoref{tab:cli_m2}. The parameters are grouped according to their relevant methods. 

\begin{table}[t!]
\centering
\begin{tabular}{|p{2.7cm}|p{5.3cm}|}
\hline
\multicolumn{2}{|c|}{\textbf{Method 1: Ansatz Benchmarking}} \\
\hline
\texttt{min\_qubits} & Minimum number of qubits \\
\texttt{max\_qubits} & Maximum number of qubits \\
\texttt{skip\_qubits} & Step size for qubit sweep \\
\texttt{max\_circuits} & Circuits per qubit size \\
\texttt{num\_layers} & Number of ansatz layers \\
\texttt{init\_state} & Initial state to encode \\
\texttt{n\_measurements} & Number of measured qubits \\
\texttt{num\_shots} & Shots per circuit execution \\
\texttt{data\_reupload} & Enable/disable data re-uploading \\
\texttt{nonoise} & Run with noiseless simulator \\
\hline
\end{tabular}
\caption{\textbf{Parameters for Method~1 (circuit benchmarking).}}
\label{tab:cli_m1}

\vspace{0.3cm}

\centering
\begin{tabular}{|p{3.7cm}|p{4.3cm}|}
\hline
\multicolumn{2}{|c|}{\textbf{Method 2: QRL Training Loop}} \\
\hline
\texttt{num\_layers} & Number of ansatz layers \\
\texttt{n\_measurements} & Number of measured qubits \\
\texttt{num\_shots} & Shots per circuit execution \\
\texttt{data\_reupload} & Enable/disable data re-uploading \\
\texttt{total\_steps} & Maximum total environment steps \\
\texttt{learning\_start} & Steps before updates begin training the network\\
\texttt{params\_update} & Interval between parameter updates \\
\texttt{target\_update} & Interval between target updates \\
\texttt{batch\_size} & Replay buffer batch size \\
\texttt{exploration\_fraction} & Fraction of steps used for exploration \\
\texttt{tau} & Soft update factor \\
\texttt{nonoise} & Run with noiseless simulator \\
\hline
\end{tabular}
\caption{\textbf{Parameters for Method~2 (QRL training loop).}}
\label{tab:cli_m2}
\end{table}

\subsubsection{Parameterized Quantum Circuit Ansatz}
\label{app:ansatz}

The ansatz encodes states and outputs Q-values for actions. Its structure is as follows (also refer ~\autoref{fig:PQC} for details):
\begin{itemize}
  \item \textbf{Width and measurements:} For method 1 the ansatz width and measured qubits are determined by parameters set by the user. For method 2 the width of the circuit is set by the greater of the observation space size and the action space size of the FrozenLake map chosen. The total number of qubits that are measured reflect the size of the action space of the FrozenLake map. 
  \item \textbf{Encoding:}  
  The environment provides the state of the agent by providing an integer ID of the tile on which the agent is. We convert the state into its binary representation, and each state bit is encoded with an $R_x$ gate if that bit equals $1$, if the bit is $0$, no rotation gate is applied.
  \item \textbf{Variational layers:}  
  Each layer has $R_y$ and $R_z$ rotations plus $CZ$ entangling gates. The number of layers is set by \texttt{num\_layers}.
  \item \textbf{Data re-uploading:}  
  Optional extra $R_x$ gates per layer (\texttt{data\_reupload}) improve expressivity and performance of Q-DQN~\cite{perez2020data, Barthe2024gradientsfrequency}.
  \item \textbf{Shots and noise:}  
  The number of shots per execution is set by \texttt{num\_shots}. Setting the \texttt{nonoise} flag uses an ideal simulator.
\end{itemize}

\subsubsection{Training Workflow and Optimizers}
\label{app:training}

\begin{figure}[t!]
    \centering
    \includegraphics[width=\linewidth]{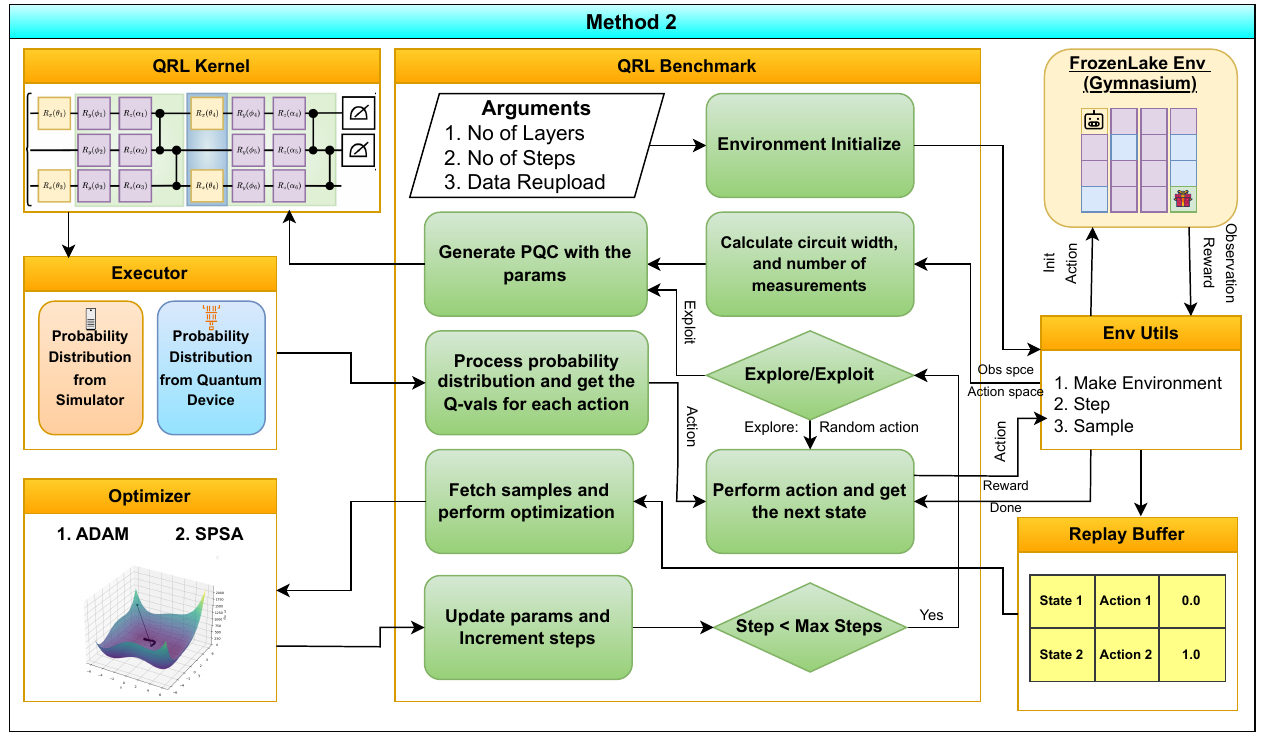}
    \caption{\textbf{Method 2 Detailed Overview}: This figure provides a detailed overview of all the important steps for Method 2 of QRL Benchmark.}
    \label{fig:M2_detailed}
\end{figure}

The training workflow is described in detail in ~\autoref{fig:M2_detailed}. Method 2 runs a Quantum Deep Q-Network (Q-DQN)in. Actions are chosen by taking the maximum Q-value from the expectation value $\langle Z \rangle$ on the measured qubits. We employ a $\epsilon$-greedy schedule \cite{dann2022guarantees} to manage the exploration/exploitation schedule. Initially, nearly all the steps are exploration and then probability of exploration keeps decaying over a fraction of the total steps. 

Training uses a replay buffer. After \texttt{learning\_start} steps, batches of size \texttt{batch\_size} are sampled to update parameters. Updates happen every \texttt{params\_update} steps. The target network is updated every \texttt{target\_update} steps, with optional soft updates using \texttt{tau}.  

Optimizers include:
\begin{itemize}
  \item \textbf{SPSA:} noise robust and requires fewer circuit evaluations,
  \item \textbf{ADAM:} adaptive, gradient-based.
\end{itemize}
The modular API supports adding new optimizers easily.

\subsubsection{Performance Metrics}
\label{app:metrics}

\textbf{Method 1:} Metrics include:
\begin{itemize}
  \item Hellinger and normalized fidelity,
  \item algorithmic and normalized depth,
  \item quantum and wall-clock runtime,
  \item volumetric plots (fidelity vs.\ depth/width).
\end{itemize}

\textbf{Method 2:} Metrics include:
\begin{itemize}
  \item \textbf{Per-step:} circuit evaluations, exploration/exploitation counts, environment interactions.
  \item \textbf{Aggregate:} total steps, gradient evaluations, episodes completed, success rate, average return per episode and run.
  \item \textbf{Timing:} total and per-step time for quantum execution, environment interaction, and gradient evaluation.
\end{itemize}
All results are printed to the console and plotted for analysis.

\end{document}